\documentstyle[preprint,aps,eqsecnum]{revtex}

\def\beq#1{\begin{equation} \label{#1}}
\def\eeq{\end{equation}}
\newcommand{\bea}{\begin{eqnarray}}
\newcommand{\eea}{\end{eqnarray}}
\def\bra#1{\left\langle #1\right\vert}
\def\ket#1{\left\vert #1\right\rangle}
\def\epsp{\epsilon^{\prime}}
\def\NPB{{ Nucl. Phys.} B}
\def\PLB{{ Phys. Lett.} B}
\def\PRL{ Phys. Rev. Lett.}
\def\PRD{{ Phys. Rev.} D}

\begin{document}
{
\tighten

\title {The GSI method for studying neutrino mass differences \\
 For Pedestrians}
\author{Harry J. Lipkin\,\thanks{Supported in part by
U.S.
Department of Energy, Office of Nuclear Physics, under contract
number
DE-AC02-06CH11357.}}
\address{ \vbox{\vskip 0.truecm}
  Department of Particle Physics
  Weizmann Institute of Science, Rehovot 76100, Israel \\
\vbox{\vskip 0.truecm}
School of Physics and Astronomy,
Raymond and Beverly Sackler Faculty of Exact Sciences,
Tel Aviv University, Tel Aviv, Israel  \\
\vbox{\vskip 0.truecm}
Physics Division, Argonne National Laboratory
Argonne, IL 60439-4815, USA\\
~\\harry.lipkin@weizmann.ac.il
\\~\\
}

\maketitle

\begin{abstract}

New experiment studying radioactive ion before K-capture weak decay
provides new method for investigating creation of coherent final state
producing $\nu$ oscillations. Single radioactive ion passes through
known electromagnetic interactions in storage ring before decaying by first
order weak interaction conserving momentum described by Fermi's golden rule.
Initial state must contain
coherent linear combinations of two states with
same momentum difference producing final state oscillations.
Following passage of ion through storage ring provides information
about $\nu$  masses and mixing without detecting
$\nu$. Observing every weak decay avoids suppression in conventional
oscillation experiments by low $\nu$ absorption cross sections.
Normally unobservable long wave lengths made observable by
long distance circulation around storage ring.
No oscillations can be observed in ``missing mass" experiment where both energy and momentum are conserved. Energy-time uncertainty allows decay into same final state of two initial state components with same momentum difference and
slightly different energies. Relative phase of these
two components changes with time and can produce
oscillations between Dicke superradiant and subradiant states. Analysis of
oscillations in simple model with no fudge factors gives value for the
squared neutrino mass difference in the same ball park as the KAMLAND value.
Treatment consistent with quantum mechanics and causality.

\end{abstract}

} 


\def\beq#1{\begin{equation} \label{#1}}
\def\eeq{\end{equation}}
\def\bra#1{\left\langle #1\right\vert}
\def\ket#1{\left\vert #1\right\rangle}
\def\epsp{\epsilon^{\prime}}
\def\NPB{{ Nucl. Phys.} B}
\def\PLB{{ Phys. Lett.} B}
\def\PRL{ Phys. Rev. Lett.}
\def\PRD{{ Phys. Rev.} D}

\section {Introduction - The basic paradox of neutrino oscillations}

\subsection {Interference is possible only if we can't know everything}

Neutrino oscillations are produced from a coherent
mixture of different  $\nu$ mass eigenstates. If the $\nu$ is
produced in a reaction where all other particles have definite momentum and
energy the $\nu$ mass is determined
by conservation of energy and momentum. Interference between amplitudes from different $\nu$
mass eigenstates is not observable in such a ``missing mass"  experiment.
Something must prevent knowing the neutrino mass from
conservation laws. Ignorance alone does not produce interference. Quantum
mechanics must hide information. To check how coherence and oscillations can occur we investigate what is known and what information is hidden by
quantum mechanics.

A recent GSI experiment\cite{gsi} describes a unique opportunity for this
investigation. Oscillations are
observed in the decay time of a radioactive ion circulating in a storage ring.
The decay by atomic electron capture emits an unobserved monoenergetic
neutrino. This oscillation offers a new and very interesting method for
determining neutrino masses and mixing angles\cite{gsikienle,gsifaber,gsihjl}.
The detailed energy-momentum analysis is much simpler here than in other oscillation
experiments. The initial radioactive ``Mother" ion is in a one-particle
state with a definite mass moving in a storage ring. There is no
entanglement\cite{Zoltan} since no other particles are present.

The search for what is known and what is hidden
leads to energy-time uncertainty.  The
time interval between the last time the initial ion was observed without
decay and the observed decay time is sufficiently
short to allow enough violation of energy conservation to prevent its use in a missing
mass experiment. This is related to the line broadening of
a decay observed in a time short compared to its natural line width.
The decay to two final states is described by two Breit-Wigner energy
distributions. These are separated at long times. But in the GSI experiment\cite{gsi}  the
decay time is  sufficiently short to make the separation negligible in
comparison with their broadened widths.  The transition can occur coherently from two components of the
initial state with different
energies and momenta into a same final state with a different
common energy and the same momentum difference. The sum of the transition amplitudes from
these two components of an initial state to the same final state
depends on their relative phase. Changes in this phase can produce oscillations.

\subsection {Energy and momentum considerations}

The energy-time uncertainty is not covariant and defined only in the
laboratory system. Covariant descriptions and transformations from the
laboratory to any center-of-mass
system are not valid for a description of neutrino oscillations.

Macroscopic neutrino detectors in thermal
equilibrium with their environment destroy all interference
between states having different energies\cite{leofest}. The observation of
reactor neutrino oscillations  shows that the emitted neutrino wave function must contain
coherent linear  combinations of three mass eigenstates
with the same energy, a definite relative phase and a momentum differences
defined by the energy and mass differences.

  The final state is  a ``Daughter" ion and a $\nu_e$ neutrino, a
linear combination of several $\nu$ mass  eigenstates.  This $\nu_e$  is
a complicated wave packet
containing components with three different masses and a continuous spectrum of
energies and momenta. It would oscillate with  distance from the source in the
same way that reactor neutrinos oscillate.

We now summarize what is known and what is hidden by quantum mechanics.

\begin{enumerate}

\item The final state state has coherent pairs of states with same energy but
has neutrinos with different masses and different momenta.

\item The initial state is a one-particle state with a
definite mass.

\item The initial state can contain coherent pairs with the same momentum
difference present in the final state state but these must have different energies.

\item Momentum is conserved in the transition.

\item Energy-time uncertainty hides information and prevents use of energy
conservation.

\item The same final energy eigenstate can be produced from states
with different energies.

\item  The transition can occur coherently from two components of the
initial state with different
energies and momenta into a same final state with a different
common energy and the same momentum difference.

\item The relative phase between components with different energies changes
during the passage of the ion through the storage ring.

\item These phase changes can produce oscillations.

\end{enumerate}

A treatment of neutrino oscillations without explicit violation of
energy conservation describes a missing mass
experiment where no neutrino oscillations of any kind are allowed.
\subsection{The two principal difficulties of neutrino experiments}

\begin{enumerate}
\item Ordinary neutrino oscillation experiments are difficult because
\begin{itemize}
 \item The tiny neutrino absorption cross section makes number of neutrino
 events actually used many orders of magnitude smaller than
 number of undetected neutrinos.
 \item The oscillation wave lengths are so large that it is difficult to
actually follow even one oscillation period in any experiment.
\end{itemize}
\item
This experiment opens up a new line for
dealing with these difficulties
\begin{itemize}
\item The oscillation is measured without detecting the $\nu$.
Detection of every $\nu$ creation event
avoids the losses from the low neutrino absorption cross section.
   \item The long wave length problem is solved when
ions move a long distance circulating around in a storage
ring and show many oscillations in same experiment.

\end{itemize}
\end{enumerate}

\section {The basic physics of the GSI experiment}

\subsection {A first order weak transition}

The initial state wave function  $\ket {i(t)}$ is a ``Mother" ion wave packet containing components with different momenta. Its development in time is described by an unperturbed
Hamiltonian denoted by $H_o$ which
describes the motion of the initial and final states in the electromagnetic
fields constraining their motion in a storage ring.
\beq{timedep}
\ket{i(t)} = e^{iH_ot} \ket {i(0)}
\eeq

The time $t = 0$ is defined as the time
of entry into the apparatus.
Relative phases of wave function components with different momenta are determined by localization in space at the point of entry into the apparatus. Since plane waves have equal
amplitudes over all space, these relative phases are seriously constrained by
requiring that the probability of finding the ion outside the storage ring must be zero.

A first-order weak decay is described by the Fermi Golden
Rule.
The  transition probability per unit time at time $t$ from an initial state
$\ket{i(t)}$ to a final state $\ket{f}$ is

\beq {fermi}
W(t) = {{2\pi}\over{\hbar}}|\bra{f} T \ket {i(t)}|^2\rho(E_f) =
{{2\pi}\over{\hbar}}|\bra{f} T e^{iH_ot} \ket {i(0)}|^2\rho(E_f)
\eeq
where $T$ is the transition operator and $\rho(E_f)$ is the density of final states  .
The transition operator $T$ conserves momentum.

If two components of the initial state with slightly different energies can
both decay into the same final state, their relative
phase changes linearly with time and can produce changes in the
transition matrix element.
The quantitative result and the question of whether oscillations
can be observed  depend upon the evolution of the
initial state.
The neutrino is not detected in the GSI experiment\cite{gsi}, but the information
that a particular linear combination of mass and momentum eigenstates would
be created existed in the system.
 Thus the same final state
can be created by either of three initial states that have the same momentum
difference.
Violation of energy conservation allows the decay
and provides a new method for  investigating the creation of such a coherent
state.



\subsection {Time dependence and internal clocks}

A measurement of the time between  the
initial observation and the decay of a radioactive ion circulating in a storage
ring depends upon the existence of an internal clock in the
system.

\begin {enumerate}

\item An initial ion in a one-particle energy eigenstate has no
clock. Its propagation in time is completely described by a single unobservable
phase.

\item If the initial ion is in a coherent superposition of different energy
 eigenstates, the relative phase of any pair changes with energy. This phase
defines a clock which can measure the time between initial observation and
decay.

\item If the decay transition conserves energy, the final states produced by
the transition must also have different energies.

\item The decay probability  is proportional to the square of the sum of  the
transition matrix elements to all final states. There are no interference terms
between final states with different energies and their relative phases are
unobservable.

\end{enumerate}

The probability $P_i(t)$ that the ion is still in its initial state at time $t$
and not yet decayed satisfies an easily solved differential equation,
\beq{nonexp1}
\frac {d}{dt} P_i(t) = - W(t)
 P_i(t)
; ~ ~ ~ ~ ~
\frac {d}{dt} log (P_i) = - W(t)
; ~ ~ ~ ~ ~ P_i(t) = e^{-\int W(t)dt}
\eeq

If $W(t)$ is independent of
time eq. (\ref{nonexp1}) gives an exponential decay.
The observation of a nonexponential decay implies that $W(t)$ is time dependent.
A non-exponential decay can occur only if there is a violation of energy
conservation. All treatments which assume energy conservation; e.g.\cite{Zoltan} will
only predict exponential decay.
Time dependence  can
arise if the initial ion is in a coherent superposition of different energy
eigenstates, whose relative phases change with time. This phase defines a clock
which can measure the time between initial observation and decay. Since the
time $dt$ is infinitesimal, energy need not be conserved in this transition.

$W(t)$
depends upon the unperturbed propagation of the initial state before
the time $t$ where its motion in the storage ring is
described by classical electrodynamics.
Any departure from exponential decay must come from the evolution in time of
the initial unperturbed state. This
can change the wave function at the time of the decay and therefore the value of
the transition matrix element. What happens after the decay cannot change the
wave function before the decay. Whether or not and how the final neutrino
is detected cannot change the decay rate.

\subsection {The role of Dicke superradiance}

Dicke\cite{Super} has shown that whenever two initial state components can
produce amplitudes for decay into the same final state, a linear combination
called ``superradiant" has both components interfering
constructively to enhance the transition.
The orthogonal state  called ``subradiant" has maximum destructive interference and
may even produce a cancelation.

The wave function of the initial state before the transition can contain pairs
of components with a momentum difference allowing both to decay into the
same final state.  This wave function can be expressed as a linear combination of superradiant and  subradiant states with a relative magnitude that
changes with time. The variation between superradiant and subradiant wave
functions affects the transition matrix element and can give rise to oscillations
in the decay probability. Since the momentum difference depends on the mass difference
between the two neutrino eigenstates these oscillations  can provide information about neutrino masses.

\section{Detailed analysis of a simplified model}

\subsection {The initial and final states for the transition matrix}

The final state is a daughter ion and an electron neutrino.

\begin{enumerate}
\item{The neutrino wave packet contains different masses, energies and momenta}

\item{Oscillation experiments use macroscopic detectors in thermal
equilibrium which kill coherence between states with different energies}

\item{Observed oscillations arise only from components with same energy,
different masses and different momenta}

\end{enumerate}

The initial state before the transition is a mother ion wave packet.

\begin{enumerate}

\item{Need to know evolution of wave packet during passage around
the storage ring}

\item {Not easily calculated. Requires knowing path through straight sections, bending sections and focusing electric and magnetic fields}

\end{enumerate}.

\subsection{Dicke superradiance and subradiance in the experiment}

Consider the transition from a simplified initial state for the
``mother" ion  with only two components denoted by $\ket{\vec P}$ and $\ket{\vec P +
\delta \vec P}$ having momenta $\vec P$ and $\vec P +\delta \vec P$ with energies $E$ and $E +
\delta E$. These can decay coherently into a final state which
produces the observed oscillations. The final state has a recoil
``daughter" ion and an electron neutrino which is a linear  combination of  two
neutrino mass eigenstates.  The
oscillations must be produced in  a final state having two components
with the same energy. The recoil ``daughter" ion   in both components nust have
the same momentum $\vec P_R$ and  energy $E_R$ in order that the two final state
components be coherent.

The final state $\ket{f(E_\nu)}$ has an electron neutrino with energy $E_\nu$.
\beq{final2com}
\ket{f(E_\nu)}\equiv \ket{\vec P_R;\nu_e(E_\nu)}  =
\ket{\vec P_R;\nu_1(E_\nu)}\bra{\nu_1}\nu_e\rangle + \ket{\vec P_R;\nu_2(E_\nu)}\bra{\nu_2}\nu_e\rangle
\eeq
where $\nu_1$ and $\nu_2$ denote neutrino mass eigenstates with masses $m_1$ and
$m_2$.  $\bra{\nu_1}\nu_e\rangle$ and $\bra{\nu_2}\nu_e\rangle$ are elements of the neutrino mass mixing matrix, commonly expressed in terms of a mixing angle denoted by
$\theta$.
\beq{final3com}
\cos \theta \equiv \bra{\nu_1}\nu_e\rangle; ~ ~ ~
 \sin \theta \equiv \bra{\nu_2}\nu_e\rangle; ~ ~ ~\ket{f(E_\nu)}
= \cos \theta \ket{\vec P_R;\nu_1(E_\nu)}+ \sin \theta \ket{\vec P_R;\nu_2(E_\nu)}
\eeq
After a very short time two components with different initial
state energies can decay into a final state which has two components with the
same energy and a
neutrino state having two components with the same momentum difference
$\delta \vec P$ present in the initial state.

The momentum conserving transition matrix elements between the two initial
momentum components to final states with the same energy and momentum difference
$\delta \vec P$ are
\beq{transcom}
\bra{f(E_\nu)} T \ket {\vec P)} = \cos \theta \bra {\vec P_R;\nu_1(E_\nu)}T \ket {\vec P)}
;~ ~ ~
\bra{f(E_\nu)} T \ket {\vec P + \delta \vec P)} =\sin \theta \bra {\vec P_R;\nu_2(E_\nu)}T
\ket {\vec P + \delta \vec P)}
\eeq
We neglect transverse momenta and set
$\vec P\cdot\vec p_\nu \approx P p_\nu$ where $P$ and $p_\nu$ denote the
components of the momenta in the direction of the incident beam.
The  Dicke superradiance analog \cite{Super} is seen by defining superradiant and
subradiant states.
\beq{super}
\ket{Sup(E_\nu)}\equiv
\cos \theta \ket {P)} + \sin \theta \ket {P + \delta P)}; ~ ~ ~
\ket{Sub(E_\nu)}\equiv \cos \theta \ket {P + \delta P)}- \sin \theta \ket {P)}
\eeq
The transition matrix elements for these two states are then
\beq{trans}
\frac{\bra {f(E_\nu)} T \ket {Sup(E_\nu)}}{\bra{f} T \ket {P }} =[\cos \theta +
\sin \theta ]
; ~ ~ ~
\frac{\bra {f(E_\nu)} T \ket {Sub(E_\nu)}}{\bra{f} T \ket {P }} =  [\cos \theta -
\sin \theta ]
\eeq
where we have neglected the dependence of the transition operator $T$ on the
small change in the momentum $P$.
The squares of the transition matrix elements are

\beq{transsupsubsq}
\frac{|\bra {f(E_\nu)} T \ket {Sup(E_\nu)}|^2}{|\bra{f} T \ket {P }|^2} =
[1 + \sin 2 \theta ]
; ~ ~ ~
\frac{|\bra {f(E_\nu)} T \ket {Sub(E_\nu)}|^2}{ |\bra{f} T \ket {P }|^2 }=
[1 - \sin 2 \theta ]
\eeq

For maximum neutrino mass mixing, $\sin 2 \theta =1$ and
\beq{transsupsubmax}
|\bra {f(E_\nu)} T \ket {Sup(E_\nu)}|^2 =
2 |\bra{f} T \ket {P }|^2
; ~ ~ ~
|\bra {f(E_\nu)} T \ket {Sub(E_\nu)}|^2 = 0
\eeq

This is the standard Dicke superradiance in which all the transition strength
goes into the  superradiant state and there is no transition from the
subradiant state.

\subsection{Kinematics for a simplified two-component initial state.}

We first consider the transition for
each component of the wave  packet which has a momentum $\vec P$ and energy $E$
in the initial state.  The final state has a recoil ion with momentum denoted
by $\vec P_R$ and  energy $E_R$ and a neutrino with mass $m$, energy $E_\nu$
and momentum  $\vec p_\nu$. If both energy and momenta are conserved,

\beq{epcons} E_R= E - E_\nu;  ~ ~ \vec P_R = \vec P - \vec p_\nu ; ~ ~
M^2 + m^2 - M_R^2 =2EE_\nu - 2\vec P\cdot\vec p_\nu
\eeq
We again neglect transverse momenta and consider the simplified two-component initial state for the ``mother" ion having
momenta $P$ and $P + \delta P$ with energies $E$ and $E + \delta E$.
The final state has two components
having neutrino momenta $p_\nu$ and $p_\nu + \delta p_\nu$ with energies $E_\nu$ and $E_\nu + \delta E_\nu$ together with a recoil ion having the same momentum and energy for both components.
The changes in these
variables required to
produce a small change $\Delta (m^2)$ in the
squared neutrino mass are seen from eq. (\ref{epcons}) to satisfy the relation
\beq{delm3}
\frac{\Delta (m^2)}{2} =
 E \delta E_\nu
+ E_\nu \delta E -
P \delta p_\nu -p_\nu \delta P =
- E \delta E \cdot \left[1 - \frac {\delta E_\nu}{\delta E}+
\frac {p_\nu}{P} - \frac {E_\nu}{E}\right] \approx  - E\delta E
\eeq
where we have noted that the two final neutrino components have the same energy; i.e.
$\delta E_\nu = 0$, momentum conservation in the transition requires
$P \delta p_\nu = P\delta P = E \delta E$,
$E$ and $P$ are of the order of the mass $M$ of the ion
and $p_\nu$ and $E_\nu$ are much less than $M$.



The relative phase $\delta \phi$ at a time t between the two states
$\ket{P}$ and $\ket{P + \delta P}$ is given by $\delta E \cdot t$.
Equation (\ref{delm3}) relates $\delta E$ to the difference
between the squared masses of the two neutrino mass eigenstates. Thus
\beq{delphipotalt}
E\cdot \delta E =-{{\Delta (m^2)}\over{2}}; ~ ~ ~ ~
\delta \phi \approx -\delta E\cdot t =
-{{\Delta (m^2)}\over{2E}}\cdot t
= - {{\Delta (m^2)}\over{2\gamma M}}\cdot t
\eeq
where $\gamma$ denotes the Lorentz factor $E/M$.

Thus from eq.
(\ref{super}) the initial state at time t varies periodically between the
superradiant and
subradiant states.
The period of oscillation $\delta t$ is obtained by setting  $\delta \phi \approx -2\pi$,
\beq{deltat}
\delta t \approx  {{4 \pi \gamma M}\over{\Delta (m^2)}}; ~ ~ ~
\Delta (m^2) = {{4 \pi \gamma M}\over{\delta t}} \approx
2.75 \Delta (m^2)_{exp}
\eeq
where the values of $\delta t$ and $\Delta (m^2)_{exp}$ are obtained from the GSI experiment and neutrino
oscillation experiments\cite{gsikienle}.

The theoretical value (\ref{deltat}) obtained with minimum
assumptions and no fudge factors is in the same ball park as the experimental
value obtained from completely different experiments. Better values obtained
from better calculations can be very useful in determining the masses and
mixing angles for neutrinos.

\subsection{A tiny energy scale}

The experimental result sets a scale  in time  of seven
seconds. This gives a tiny energy scale for the difference between two waves
which beat with a period of seven seconds.
\beq{beat}
\Delta E \approx 2\pi \cdot \frac {\hbar}{7}
= 2\pi \cdot \frac {6.6\cdot 10^{-16}}{7} \approx 0.6\cdot 10^{-15}{\rm eV}
\eeq

This tiny energy scale must be predictable from standard quantum mechanics
using a scale from another input. The only other input available is in the propagation of the initial state through the
storage ring during the time before
the decay. One tiny scale available in the parameters that describe this
experiment is the mass-squared difference between two neutrino mass
eigenstates. This gives a prediction (\ref{deltat}) which differs from the exact value by less than a factor of three in the simplest approximation.

That these two tiny energy scales obtained from completely different inputs are
within an order of magnitude of one another suggests some relation obtainable
by a serious quantum-mechanical calculation. We have shown here that the simplest
model relating these two tiny mass scales gives a result that
differs by only by a factor of less than three.

Many other possible mechanisms might produce oscillations. The
experimenters\cite{gsi} claim that they have investigated all
of them. These other mechanisms generally involve energy scales very different from
the scale producing a seven second period.

The observed oscillation is seen to arise from the relative phase between two components of the initial
wave function with a tiny energy difference (\ref{beat}). These components travel through the electromagnetic fields required to maintain a stable orbit. The effect in these fields on the relative phase depends on the energy difference between the two components. Since the energy difference is so tiny the effect on the phase is expected to be also tiny and calculable.

\subsection{Effects of spatial dependence}

The initial wave function travels through space as well as time.
In a storage ring the ion moves through straight sections, bending sections and
focusing fields.  All must be included to obtain a reliable
estimate for $\Delta (m^2)$. That this requires a detailed complicated calculation
is seen in examining two extreme cases
\begin{enumerate}

\item Circular motion in constant magnetic field.
The cyclotron frequency is independent of the momentum of the ion.
Only the time dependent term contributes to the phase and $\delta \phi^{cyc}$ is
given by eq. (\ref{delphipotalt})

\item Straight line motion with velocity $v = (P/E)\cdot t$.
The phase of the initial state at point $x$ in space and time $t$,
its change with energy and momentum changes $\delta P$ and $\delta E$
are
\beq{phisl} \phi^{SL} = P\cdot x -E\cdot t; ~ ~ ~ ~ \delta \phi^{SL} = (\delta P\cdot v -\delta E)\cdot t= \frac{P\delta P -E\delta E}{E}\cdot t=0
\eeq
\end{enumerate}

The large difference between the two results (\ref{delphipotalt} and (\ref{phisl}) indicate that a precise determination of the details of the motion of the mother ion in the storage ring is needed before
precise predictions of the squared neutrino mass difference can be made.

\section{Conclusions}

A new oscillation  phenomenon providing information about neutrino mixing is
obtained by following the initial radioactive ion  before the decay.
Difficulties introduced in conventional $\nu$ experiments by tiny
neutrino absorption cross sections and very long oscillation wave lengths
are avoided. Measuring the decay time enables every $\nu$ event to be
observed and counted without the necessity of observing the $\nu$ via the
tiny absorption cross section. The confinement of the initial ion in a storage
ring enables long wave lengths to be measured within the laboratory.

\section{Acknowledgement}
The theoretical analysis in this paper was motivated by discussions with  Paul
Kienle at a very early stage of the experiment in trying to understand whether
the effect was real or just an experimental error.
It is a pleasure to thank him for calling my attention to this problem
at the Yukawa Institute for Theoretical Physics at Kyoto  University, where
this work was initiated during the YKIS2006 on ``New  Frontiers on QCD".
Discussions on possible experiments with Fritz Bosch, Walter Henning, Yuri
Litvinov and Andrei Ivanov are also gratefully acknowledged along with a
critical review of the present manuscript. The author also acknowledges further
discussions on neutrino oscillations as ``which path" experiments with  Eyal
Buks, Avraham Gal, Terry Goldman, Maury Goodman,  Yuval Grossman, Moty Heiblum, Yoseph Imry,
Boris Kayser, Lev Okun, Gilad Perez, Murray Peshkin, David Sprinzak, Ady Stern,
Leo  Stodolsky and Lincoln Wolfenstein.
%
\catcode`\@=11 
\def\references{
\ifpreprintsty \vskip 10ex
%
\hbox to\hsize{\hss \large \refname \hss }\else
\vskip 24pt \hrule width\hsize \relax \vskip 1.6cm \fi \list
{\@biblabel {\arabic {enumiv}}}
{\labelwidth \WidestRefLabelThusFar \labelsep 4pt \leftmargin \labelwidth
\advance \leftmargin \labelsep \ifdim \baselinestretch pt>1 pt
\parsep 4pt\relax \else \parsep 0pt\relax \fi \itemsep \parsep \usecounter
{enumiv}\let \p@enumiv \@empty \def \theenumiv {\arabic {enumiv}}}
\let \newblock \relax \sloppy
 \clubpenalty 4000\widowpenalty 4000 \sfcode `\.=1000\relax \ifpreprintsty
\else \small \fi}
\catcode`\@=12 
{\tighten

}
\end{document}